\documentclass[twocolumn,twoside,floatfix,slac_two]{revtex4}
\usepackage{graphicx}
\usepackage{fancyhdr}
\setcitestyle{square,numbers}
\begin{document}

\title{Behaviour of the EAS Age Parameter in the Knee Energy Region}

%

\author{R.K. Dey}
\affiliation{Dept of Physics, Univ. North Bengal, Siliguri, WB 734013 India}
\author{A. Bhadra}
\affiliation{High Energy \& Cosmic Ray Research Ctr, Univ. North Bengal, Siliguri, WB 734013 India}

\author{J.N. Capdevielle}
\affiliation{APC, Univ.Paris Diderot, 10 rue A.Domon, 75205 Paris, France}

\begin{abstract}
Analyzing simulated EAS events generated with the CORSIKA code, the characteristics of lateral distribution of electrons in EAS around the knee energy region of the primary energy spectrum have been studied and compared with experimental observations. The differences between the EGS4 and the NKG output of CORSIKA in respect to electron radial density distribution have been investigated. The relation between lateral and longitudinal age parameters has been studied after introducing the notion of the local age parameter that reflects the profile of the lateral distribution of electrons in EAS. The present analysis motivates the inclusion of the lateral shower age in a multiparameter study of EAS to extract information on hadronic interactions and primary composition.

\end{abstract}

\maketitle

\thispagestyle{fancy}


\section{THE DIFFERENT AGE PARAMETERS}

The concept of shower age was introduced in cascade theory to describe the stage of development of an electromagnetic (e.m.) cascade. A synthesis summarizing the works of Greisen and Nishimura-Kamata under Approximation B of cascade development \cite{coc} in respect to shower age is the following: the longitudinal age $s_{L}$ is defined here as

\begin{equation}
s_{L}=\frac {3 t}{t+2ln(E/\epsilon_{0})}   \;,
\end{equation}

where $E_{0}$ is the  energy of the primary photon generating the cascade, $t$
 is the atmospheric (divided by the electron radiation lenth in air taken as $37.1$ g-cm$^{-2}$), $\epsilon_{0}$ being the critical energy 
of $82$ MeV. 

In this theoretical context the lateral density distribution of cascade particles given by Nishimura and Kamata can be approximated by the well known Nishimura-Kamata-Greisen (NKG) structure function, 

\begin{equation}
f(r)=C(s_{\bot})(r/r_{m})^{s_{\bot}-2}(1+r/r_{m})^{s_{\bot}-4.5} \;,
\end{equation}

where $r$ is the radial distance measured from the EAS core, $r_{m}$ is the Moliere
radius, $s_{\bot}$ is the lateral age : the normalization factor C($s_{\bot}$) is given by

\begin{equation}
C(s_{\bot})  =   \frac{\Gamma(4.5-s_{\bot})}{2\pi\Gamma(s_{\bot}) \Gamma(4.5-2s_{\bot})}  \;.
\end{equation}

implying that for the density $\rho_{NKG}(r)= N_{e} f(r)$
thanks to the properties of the Eulerian function.

The relation $s_{L}=s_{\bot}$ was initially considered to hold for pure e.m.    showers and it was admitted that the average steepness of the profile of the lateral distribution has a correlation with the longitudinal development. 

Later the 3D diffusion equations were solved again by Uchaikin and Lagutin using adjoint equations and an improvement of NKG function was proposed by modulating $r_{m}$  to  $s_{L}$ as follows:

\begin{equation}
\rho_{el}(r) = (mr_{m})^{-2}\rho_{NKG}(r/m)
\end{equation}

with $m = 0.78 - 0.21 s_{L}$.

The validity of this approach was demonstrated \cite{cap} by pointing out that the experimental distributions in EAS are steeper than the $\rho_{NKG}$ but are in better agreement with Monte Carlo calculations of Hillas.

For better estimation of shower age from the experimental distributions, one of us (Capdevielle) introduced the notion of the local age parameter (LAP) $s_{Loc}$ \cite{cap}: From two neighbouring points, $i$ and $j$, one can give a lateral age parameter for any distribution $f(x)$ (where $x={r\over r_{m}}$) that characterises the best fit by a NKG-type function in [$x_{i},x_{j}$]~:

\begin{equation}
s_{ij} = {{\ln(F_{ij} X_{ij}^{2} Y_{ij}^{4.5})} \over {\ln(X_{ij} Y_{ij})}}
\end{equation}

where $F_{ij}$ = {{$f(r_{i}$)}/{$f(r_{j}$)}},
$X_{ij}$=$r_{i}$/$r_{j}$, and $Y_{ij}$=($x_{i}$+1)/($x_{j}$+1). More generally, if $r_{i} \rightarrow r_{j}$, this suggests the definition of the LAP $s(x)$ (or $s(r)$) at each point~:

\begin{equation}
s(r) = {1 \over {2x+1}} \left( (x+1) {{\partial{\ln f}} \over {\partial{\ln x}}} + (2 + \beta_{0})x + 2 \right)
\end{equation}

If $\beta_{0}$=4.5, $f_{NKG}(r)$ with $s$=$s(r)$ can be used to fit $f$ in the neighbourhood of $r$.

The identification $s(r)= s_{ij}$ for $r=\frac{r_{i} +r_{j}}{2}$ remains valid for the experimental distributions as long as they are approximated by monotonic decreasing functions versus distance: A typical behaviour of s(r) was infered with a characterised minimum value of the parameter near $30-50$ m from the axis followed by a general increase at large distance and it suggests a relation $s_{L}\sim 1.25-1.3s_{\bot}$. 

After verification of the behaviour of the LAP through experimentally observed lateral distributions and particularly a detail study of the parameter with the Akeno data \cite{nag}, this approach was validated by the rapporteurs of the ICRC from 1981 to 1985 \cite{rap}. This procedure was also used in extension to calculate the radio effect of very large EAS~\cite{sup}.

\section{3D SIMULATION OF EAS}

In the present work, the high energy (above $80 {\rm GeV/n}$) hadronic interaction model QGSJET 01 version 1c has been used in combination with the low energy (below $80 {\rm GeV/n}$) hadronic interaction model GHEISHA (version 2002d) in the framework of the CORSIKA Monte Carlo program version 6.600 \cite{cor} to generate EAS events. The simulated shower library
produced consists mainly of $10,000$ EAS for each primary species Proton, Helium and Iron in the primary energy interval of $10^{14}$ eV to $3 \times 10^{16}$ eV.

Taking the opportunity of calculating simultaneously the e.m. component via both the options, the EGS and the NKG option (implemented following \cite{cap} by exploiting the relation (4) in the subcascades treatment) as admitted in CORSIKA package, we have carried out parallel simulations. We ascertain that when the calculation of the electron component is carried out with relation (4), the situation with the NKG inspired procedure implemented in CORSIKA is more close to the experimental data and also to the calculation with the EGS, as shown in fig.1. The NKG option gives a slighly larger density with steeper radial distribution in compare to the EGS option. A small density excess appears for pure electromagnetic cascades near the axis for the Corsika NKG option. Such an excess appears also in proton induced air showers. However, over a large band of densities between the radial distance $10m$ up to $100m$ from the axis, a tolerable agreement is noticed. 

For Fe primaries, both the options produce older profile near the axis with density excess (particularly in the case of NKG option) between $2-10m$ distance. The average energy of the positrons is quite small in the case of iron initiated showers and the cross section of positron annihilation becomes more important for the lower part of the cascade. This effect is probably enhanced by the longer path of the electrons in the geomagnetic field (and larger energy loss by ionisation). Therefore the NKG option is not a very good option for the simulation of heavy nuclei initiated inclined showers after the maximum development of cascade. Near vertical showers in low altitudes are nevertheless acceptable. Conversely for very high energy primaries the average energy of the positrons remains important and for proton and photon showers NKG option is still useful to calculate a large number of cascades in a short time.                                           

At larger distances a slight deficit in densities appears with the NKG option; it seems to come from the different treatment of the multiple coulomb scattering in the EGS and also may due to the enhanced  path of the muons in the geomagnetic field which results in greater loss of energy by ionisation and subsequently their decay produces more electrons. The EGS takes into account the photoproduction inside the e.m. subcascades. 

\begin{figure}
\centering
\includegraphics[width=0.5\textwidth,clip]{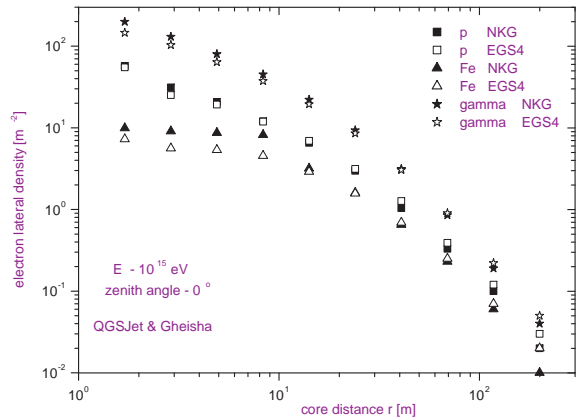} \hfill
\caption{Comparison of  radial distribution of electrons for CORSIKA NKG and EGS options for proton and Fe primaries.}
\end{figure}

Through a smaller sample of events simulated in the energy band $10^{18}-10^{20}$eV \cite{jnc}, we have also observed that the lateral distributions calculated at distances lower than $300$m from the axis with the NKG option as well as the total longitudinal development do not differ much from those obtained with the EGS. Consequently, the NKG option in CORSIKA remains useful for faster calculations at ultra high energy initiated by a nucleon or nuclei. This circumstance allows the calculation of radio effect as in \cite{sup} as well as the fluorescent component via the NKG option. The Landau Pomeranchuk Migdal effect (LPM)(not included in the NKG option) limits however the employment of the option beyond $10^{18}$eV for $\gamma$ primaries. 

The fluctuations in lateral shower age are much larger for proton initiated showers compared to those initiated by heavier primary.

\begin{figure}
\centering
\includegraphics[width=0.5\textwidth,clip]{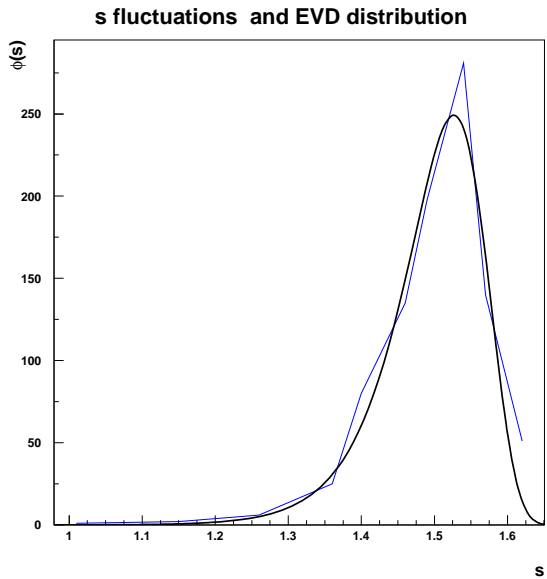} \hfill
\caption{ fluctuation  of shower age at fixed shower size $N_{e}=10^{4}$ fitted
 by an E.V.D. function $\phi(s)$ (vertical showers, sea level, p primaries)} 
\end{figure}

This distribution (fig.2) can be
 fitted by an Extreme Value Distribution (E.V.D) defined through

\begin{equation}
 \phi(s) = \frac{1}{\sigma}~
 {\rm exp}(\pm\frac{\mu-s}{\sigma}
                     - e^{\pm\frac{(\mu-s)}{\sigma}})
\label{EVD}
\end{equation}

where the parameters $\mu$ and $\sigma$ are related to the average size $\overline{s}$ and its variance $V_{s}$ by $\overline{s}~ =~ \mu \pm 0.577~\sigma$ and $V_{s}~=~1.645\sigma^{2}$ (in the case of the histogram of fig.2, $\overline{s}~=~1.495$ and $\sigma~=~0.07$).

Comparing the results of the EGS and the NKG options with Kascade data \cite{ant} in the bin $Log(N_{e}) [3.9,4.3]$, we observe that there are small discrepancies in densities even with the Lagutin NKG modified formula taking $s=1.404, N_{e}=11829$ as shown on fig.3

\begin{figure}
\centering
\includegraphics[width=0.5\textwidth,clip]{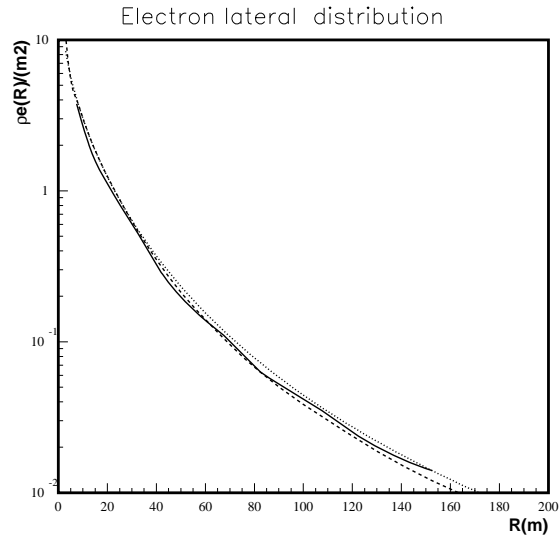} \hfill
\caption{Lateral electron distributions at Kascade level. Experimental(full line), NKG option(dashed), NKG-Lagutin modified function dotted}
\end{figure}

Those mentioned parameters (for Lagutin formula (4)) corresponds respectively to the average lateral age parameter and the average size in the size bin considered (calculated with the size spectrum in Karlsruhe).

\section{LAP AND $N_{e}$ DEPENDANCE}

The dependance of the local age parameter $s(r)$ on lateral distance exhibits the typical dependance at the level of Akeno experiment with a minimum near 30-40 m distance and larger values near at small radial distances or at large distances from the axis. An example is given  in the fig.4 for $N_{e} = 10^{6}$.

\begin{figure}
\centering
\includegraphics[width=0.5\textwidth,clip]{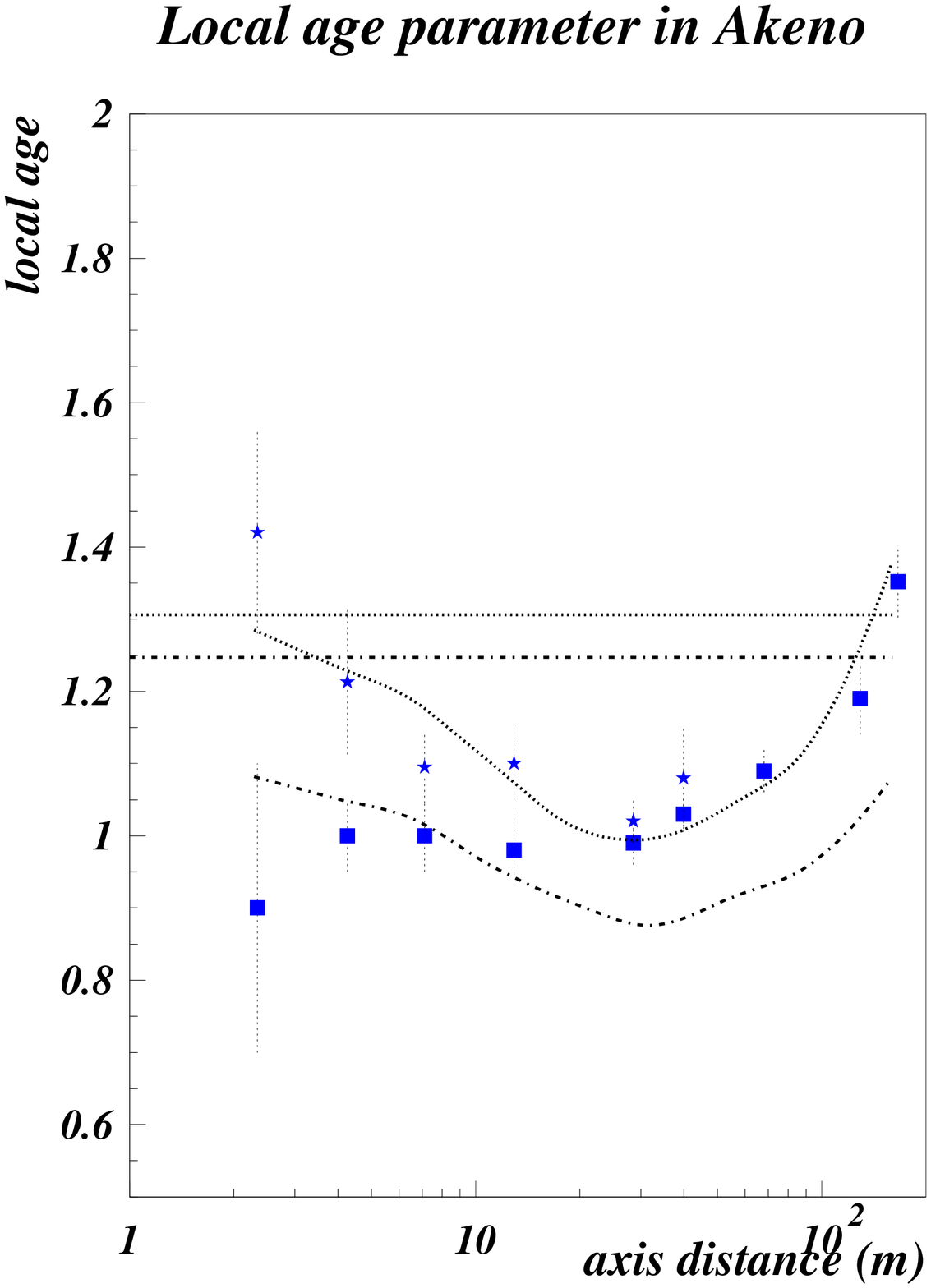} \hfill
\caption{Local age parameter versus distance at Akeno. Experimental points squares (thick scintillators), stars (thin scintillators), dotted line proton primaries, dashed line iron primary, horizontal lines longitudinal age parameter   respectively for p and iron}
\end{figure}  

The comprehensive dependance of the lateral age parameter $s$ on $N_{e}$ is shown on fig.5

\begin{figure}[t]
\centering
\includegraphics[width=0.5\textwidth,clip]{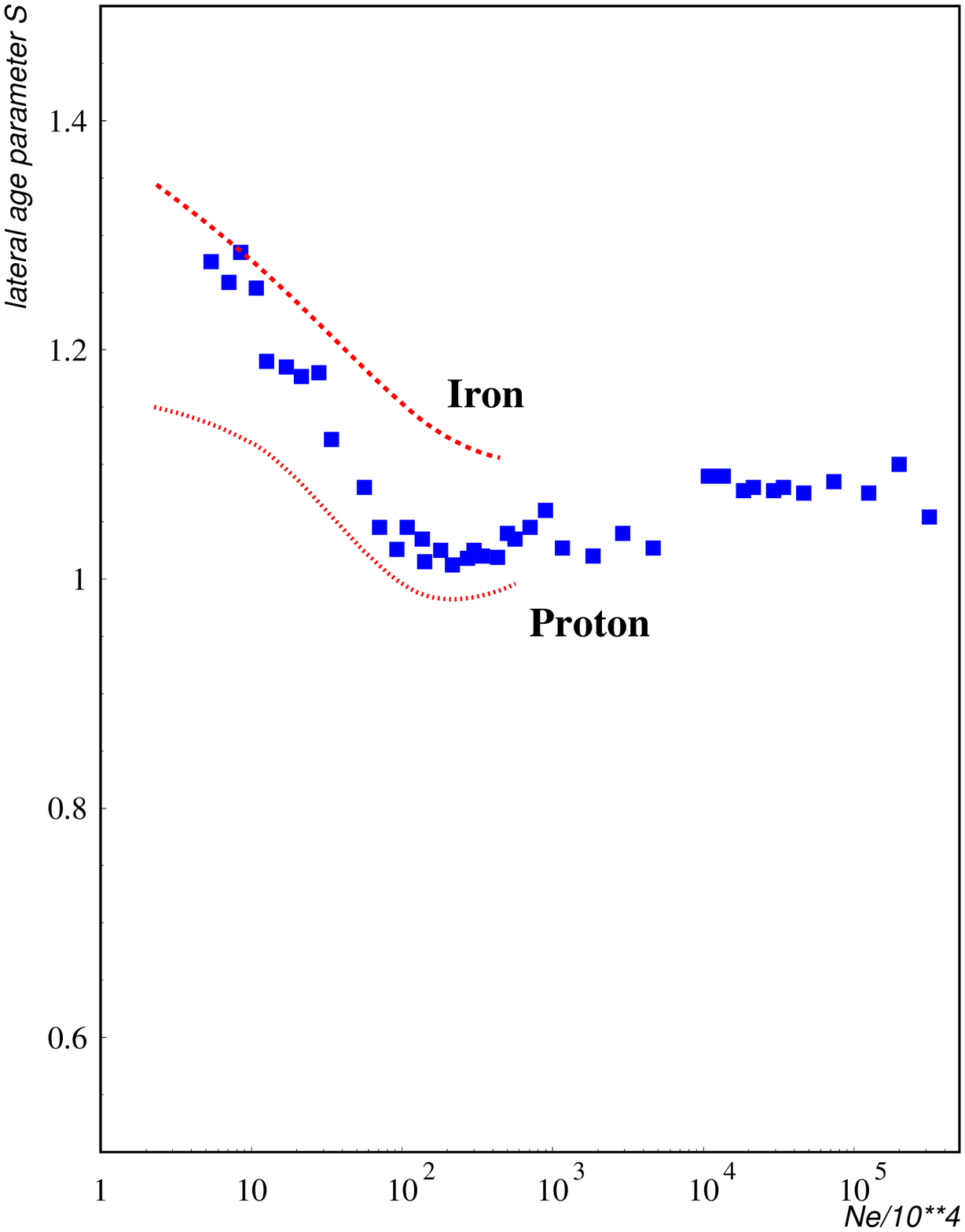} \hfill
\caption{Lateral age parameter versus size $N_{e}$  at Akeno}
\end{figure} 

The lateral age parameter is obtained as the value of the l.a.p. s(r) averaged inside the shower disc up to the  shower radius $R_{lim}$ (distance where the average lateral distribution reaches a threshold density satisfying the trigger conditions and remains detectable by the scintillator detector used i.e. for instance $1 el/m^2$ for a counter of $1 m^2$ area). One of the most simple interpretations of the minimum near the knee on fig.5 could be a mixed component progressively enriched in nuclei after the knee as suggested half a century ago by the galactic leakage,
 according to the different Larmor radius of nucleons and heavier nuclei. However,  for small energies the shower radius $R_{lim}$ is lower than $40$m whereas it exceeds $100$m at large energies above the knee; the calculation of the average $s_{\bot}$ up to those different limits may be the reason for decreasing $s$ below the knee and an increase in age beyond the knee.

\section{CONCLUSION}

 The experimental  profiles cannot be expressed in terms of NKG function with a single age parameter. Consequently, though the lateral age parameter contains information of both hadronic cascading and primary composition, its determination can be biased while fitting the experimental distributions. The characteristic dependance of local age parameter with radial distance allows a more accurate determination of $s_{\bot}$ as well as the conversion to $s_{L}$

A detailed study of dependence of  $s_{\bot}$ on shower size, included in a multiparameter analysis (muon-electron abundance, absorption length) and comparison with the Akeno and the Kascade data (in progress) should help to understand the knee region in terms of primary composition.
  
%

\begin{acknowledgments}
AB would like to thank the DST (Govt. of India) for support under the grant no. SR/S2/HEP-14/2007
\end{acknowledgments}


\bigskip 

\end{document}